\documentstyle[multicol,aps,epsfig,psfig,epsf]{revtex}

\begin{document}

\draft

\title{Permeative flows in cholesteric liquid crystals}

\author{D. Marenduzzo$^1$, E. Orlandini$^2$ and J.M. Yeomans$^1$}

\address{$^1$Department of Physics, Theoretical Physics, 1 Keble Road,
  Oxford OX1 3NP, England\\
$^2$  INFM, Dipartimento di Fisica, Universita' di Padova, Via Marzolo 8, 
35131 Padova, Italy}

\tightenlines

\maketitle

\begin{abstract}
We use lattice Boltzmann simulations to solve the Beris-Edwards equations
of motion for a cholesteric liquid crystal subjected to Poiseuille flow 
along the direction of the helical axis (permeative flow). 
The results allow us to clarify and extend the approximate analytic
treatments currently available. We find that if
the cholesteric helix is pinned at the boundaries there is an enormous
viscosity increase. If, instead, the helix is free the velocity profile is
flattened but the viscosity is essentially unchanged. We highlight the
importance of secondary flows and, for higher flow velocities, we identify
a flow-induced double twist structure in the director field
-- reminiscent of the texture characteristic of blue phases.
\end{abstract}

\begin{multicols}{2} 


Liquid crystals are fluids, typically comprising long thin
molecules, where subtle energy -- entropy balances can cause the
molecules to align to form a variety of ordered 
states\cite{degennes,chandrasekar}.  In nematic
liquid crystals the molecules tend to align parallel giving a state
with long-range orientational order.  This is usefully described by
the director field $\vec{n}$, the coarse-gained, average, molecular
orientation.  In a cholesteric or chiral nematic liquid crystal
$\vec{n}$ has a natural twist deformation in the direction perpendicular to
the molecules.  Examples of cholesteric liquid crystals are  
DNA molecules in solution, colloidal suspensions of 
bacteriophages\cite{bacterio}, and solutions of nematic mixtures such as E7 
with chiral dopants which are widely used in display devices.

Liquid crystals exhibit both an elastic and a viscous response to an
external stress.  Coupling between the director and the velocity
fields -- known as back-flow -- 
leads to strongly non-Newtonian flow behaviour.  A particularly
striking example in cholesterics is {\em permeation}.  
When a cholesteric liquid crystal is subjected to an imposed 
flow in the direction of the helix axis, its viscosity can increase 
enormously (by a factor $\sim 10^5$) when the isotropic to 
cholesteric transition is 
reached\cite{helfrich,experiments1,experiments2}.  

An explanation of permeation was given by Helfrich\cite{helfrich}.
If the director orientation is fixed in space, due for example to
anchoring effects at the wall, any flow along the helix must be linked
with a rotation of the molecules. This leads to an energy dissipation
far larger than that due to the usual molecular friction and hence a
much enhanced viscosity.  

Balancing the dissipation from the director rotation with the 
energy gained from the pressure gradient along the capillary, Helfrich
argued that the usual parabolic velocity profile is replaced by
plug-like flow, with a constant velocity across the capillary\cite{helfrich}.
To satisfy no-slip boundary conditions, however, 
the velocity must fall to zero
at the edges of the sample. De Gennes and Prost\cite{degennes} 
argue that this occurs over a length scale $\sim p$ where $p$ is the
pitch of the helix. Rey has extended these ideas to linear
and oscillatory shear\cite{rey}.
These calculations all assume small forcing, so that the
director field is not deformed by the helix. There is an
interesting suggestion in an early paper by Prost et al.\cite{jphysII}
that an increase in forcing might stabilize a modulation in the
director field in a direction perpendicular to the flow.

There are few quantitative experiments on cholesteric rheology,
mainly as it is difficult to obtain
single domain textures or a uniformly pinned 
helix\cite{experiments1,experiments2,experiments3}.
Porter et al. showed firm evidence for a high cholesteric viscosity
at low forcing\cite{experiments1}.  
This dropped to values close to those of a normal
liquid crystal as the shear rate was increased.

Helfrich's explanation of permeation is widely accepted but
many questions remain.  
There is confusion in the literature about whether the director field
must be pinned in some way to obtain a permeative flow.
It is interesting to ask whether 
distortions in the director field, induced by the flow, alter the
permeation. Does permeation persist beyond 
the regime of low forcing and
what replaces it for larger forcing ?
Finally, how does the interplay between the width of the boundary 
layer and that of the channel affect the flow?  

Given the importance of cholesteric liquid crystals in optical devices
and biological DNA solutions and the ubiquity of permeative flow in
the theory of layered liquid crystals such as cholesterics and 
smectics\cite{smectics} it is useful to develop a robust numerical method for 
solving the equations of cholesteric hydrodynamics, thus allowing 
us to probe questions that cannot be answered analytically.   
Moreover the advent of micro-channel technology means that 
quantitive experiments are likely to become feasible.

Therefore in this paper we solve numerically
the hydrodynamic equations of cholesterics and hence
clarify some of the outstanding questions about permeation.
A major conclusion is the importance of boundary conditions.
If the helix is pinned at the boundaries, e.g. by wall irregularities,
the apparent viscosity increases by orders of magnitude. 
If the boundaries are
free, the velocity profile is flat but the helical structure 
is free to drift along the flow direction and the viscosity is 
much smaller. 
As the velocity of the system is increased shear forces induce a significant
deformation of the initial helix. First the cholesteric layers
are bent into chevrons. Then at higher forcing a doubly
twisted texture is formed, the initial deformation 
being accompanied by a flow-induced twist in the perpendicular direction.

We consider the formulation of liquid crystal hydrodynamics given by
Beris and Edwards\cite{beris}.
The equations of motion are written in terms of a 
tensor order parameter ${\bf Q}$ which is related to the direction of 
individual molecules, ${\hat{n}}$, by $Q_{\alpha\beta}= \langle \hat{n}_\alpha
\hat{n}_\beta - {1\over 3} \delta_{\alpha\beta}\rangle$ where the angular
brackets denote a coarse-grained average and the Greek indices 
label the Cartesian components of ${\bf Q}$.
The tensor ${\bf Q}$ is traceless and symmetric. Its largest 
eigenvalue, $\frac {2} {3} q$, $0<q<1$, describes the 
magnitude of the order. 

The equilibrium properties of the liquid crystal are described 
by a Landau-de Gennes free energy density. This comprises
a bulk term (summation over repeated indices is implied hereafter),
\begin{eqnarray}
\frac{A_0}{2}(1 - \frac {\gamma} {3}) Q_{\alpha \beta}^2 - 
          \frac {A_0 \gamma}{3} Q_{\alpha \beta}Q_{\beta
          \gamma}Q_{\gamma \alpha} 
+ \frac {A_0 \gamma}{4} (Q_{\alpha \beta}^2)^2,
\label{eqBulkFree}
\end{eqnarray}
and a distortion term, which for cholesterics is\cite{degennes}
\begin{equation}
\frac{K}{2} \left[\left(\partial_\beta Q_{\alpha \beta}\right)^2
+ \left(\epsilon_{\alpha \zeta\delta }
\partial_{\zeta}Q_{\delta\beta} + \frac{4\pi}{p}
Q_{\alpha \beta}\right)^2 \right], 
\end{equation}
where $K$ is an elastic constant.
The tensor $\epsilon_{\alpha \zeta\delta}$ is  the Levi-Civita
antisymmetric third-rank tensor, $A_0$ is a constant and $\gamma$
controls the magnitude of order.
The anchoring of the director field on the boundary surfaces 
is ensured by adding a surface term 
proportional to $(Q_{\alpha \beta}-Q_{\alpha \beta}^0)^2$,
with $Q_{\alpha \beta}^0$ chosen in such a way 
that the director has the desired orientation at the boundaries.

The equation of motion for {\bf Q} is \cite{beris}
\begin{equation}
(\partial_t+{\vec u}\cdot{\bf \nabla}){\bf Q}-{\bf S}({\bf W},{\bf
  Q})= \Gamma {\bf H}
\label{Qevolution}
\end{equation}
where $\Gamma$ is a collective rotational diffusion constant.
The first term on the left-hand side of Eq. (\ref{Qevolution})
is the material derivative describing the usual time dependence of a
quantity advected by a fluid with velocity ${\vec u}$. This is
generalized for rod-like molecules by a second term 
\begin{eqnarray}\label{S_definition}
{\bf S}({\bf W},{\bf Q})
& = &(\xi{\bf D}+{\bf \Omega})({\bf Q}+{\bf I}/3)+ ({\bf Q}+
{\bf I}/3)(\xi{\bf D}-{\bf \Omega}) \\ \nonumber 
& - & 2\xi({\bf Q}+{\bf I}/3){\mbox{Tr}}({\bf Q}{\bf W})
\end{eqnarray}
where Tr denotes the tensorial trace, while 
${\bf D}=({\bf W}+{\bf W}^T)/2$ and
${\bf \Omega}=({\bf W}-{\bf W}^T)/2$
are the symmetric part and the anti-symmetric part respectively of the
velocity gradient tensor $W_{\alpha\beta}=\partial_\beta u_\alpha$.
The constant $\xi$ depends on the molecular
details of a given liquid crystal.
The term on the right-hand side of Eq. (\ref{Qevolution})
describes the relaxation of the order parameter towards the minimum of
the free energy. The molecular field ${\bf H}$ 
is given by
\begin{equation}
{\bf H}= -{\delta {\cal F} \over \delta {\bf Q}}+({\bf
    I}/3) Tr{\delta {\cal F} \over \delta {\bf Q}}.
\label{molecularfield}
\end{equation}   

The three-dimensional fluid velocity, $\vec u$, obeys the continuity equation
and the Navier-Stokes equation, 
\begin{eqnarray}\label{navierstokes}
\rho(\partial_t+ u_\beta \partial_\beta)
u_\alpha & = & \partial_\beta (\Pi_{\alpha\beta})+ 
\eta \partial_\beta(\partial_\alpha
u_\beta + \partial_\beta u_\alpha\\ \nonumber
& + & (1-3\partial_\rho
P_{0}) \partial_\gamma u_\gamma\delta_{\alpha\beta}),
\end{eqnarray}
where $\rho$ is the fluid density and $\eta$ is an isotropic
viscosity. The stress tensor $\Pi_{\alpha\beta}$ 
necessary to describe liquid crystal hydrodynamics is:
\begin{eqnarray}
\Pi_{\alpha\beta}= &-&P_0 \delta_{\alpha \beta} +2\xi
(Q_{\alpha\beta}+{1\over 3}\delta_{\alpha\beta})Q_{\gamma\epsilon}
H_{\gamma\epsilon}\\\nonumber
&-&\xi H_{\alpha\gamma}(Q_{\gamma\beta}+{1\over
  3}\delta_{\gamma\beta})-\xi (Q_{\alpha\gamma}+{1\over
  3}\delta_{\alpha\gamma})H_{\gamma\beta}\\ \nonumber
&-&\partial_\beta Q_{\gamma\nu} {\delta
{\cal F}\over \delta\partial_\alpha Q_{\gamma\nu}}
+Q_{\alpha \gamma} H_{\gamma \beta} -H_{\alpha
 \gamma}Q_{\gamma \beta} . 
\label{BEstress}
\end{eqnarray}
$P_0$ is a constant in the simulations reported here.

The differential equations (\ref{Qevolution}) and 
(\ref{navierstokes}) are coupled. 
Unless the flow field is zero ($\vec{u}=0$) the dynamics given by
Eq. (\ref{Qevolution}) are not purely
relaxational.
Conversely, the order parameter field affects the dynamics of the flow field 
through the stress tensor. This is the back-flow coupling.
To solve these equations we use a lattice
Boltzmann algorithm. Details and validation of using this
method to solve the Beris--Edwards model were given in Ref. 
\cite{colin,proc}.

We consider a cholesteric liquid crystal which is sandwiched 
between two plates a distance $L$ apart along the $z$-axis. The 
axis of the cholesteric helix lies in a direction parallel to the plates 
which we shall take as the $y$-axis.  The primary flow is also    
along $y$, and is imposed via a pressure gradient. 
In the steady state, however, there can be secondary flow
so that the modelling must be fully three-dimensional. 
This geometry is the one for which the permeation mode is 
expected.  The elastic constants and viscosities are taken to 
have typical values for a cholesteric and we consider channel widths 
$L \sim \mu m$. Typically, a cholesteric pitch was discretised 
by $64$ lattice points, while $10^6$ iterations were performed. 
Other details of the parameters for each simulation 
are given in the figure captions. There are no-slip velocity 
boundaries on the walls.

The results in Figure 1 aim to compare the cases where the 
director at the wall is pinned or free to rotate.
Consider first the case where the director is free. As a
bench mark we show in Figure 1a results from simulations where the
back-flow, ie the effect of the director field on the flow field, is
turned off. The flow profile is as expected the parabola of 
Poiseuille flow.  The cholesteric helix drifts at a 
slightly ($\sim 10\%$) smaller
velocity than the flow. It is slightly bent by the flow 
into a chevron structure.

The origin of the chevron pattern can be identified by considering
Eq. \ref{Qevolution}. Focussing on 
the center of the channel, ${\bf S}$ is zero, and the term 
$\vec u\cdot\vec{\nabla} {\bf Q}$ -- typical of this
flow mode, in which the director field is not constant along the
flow direction -- must be balanced by a drift of 
the layers, $\partial_t{\bf Q}$. However, due to the parabolic shape of the 
velocity field, these cannot cancel exactly and it is necessary
to allow for a non-zero molecular field, resulting in the
observed bending. The dominant elastic deformations associated with
this steady state solution are splay--bend.
The director field also develops a small component along the
flow direction. This is caused by the
shear forces contained in the tensor ${\bf S}$, which is non-zero 
away from the center of the channel.

\begin{figure}[hbtp]
\centerline{\psfig{figure=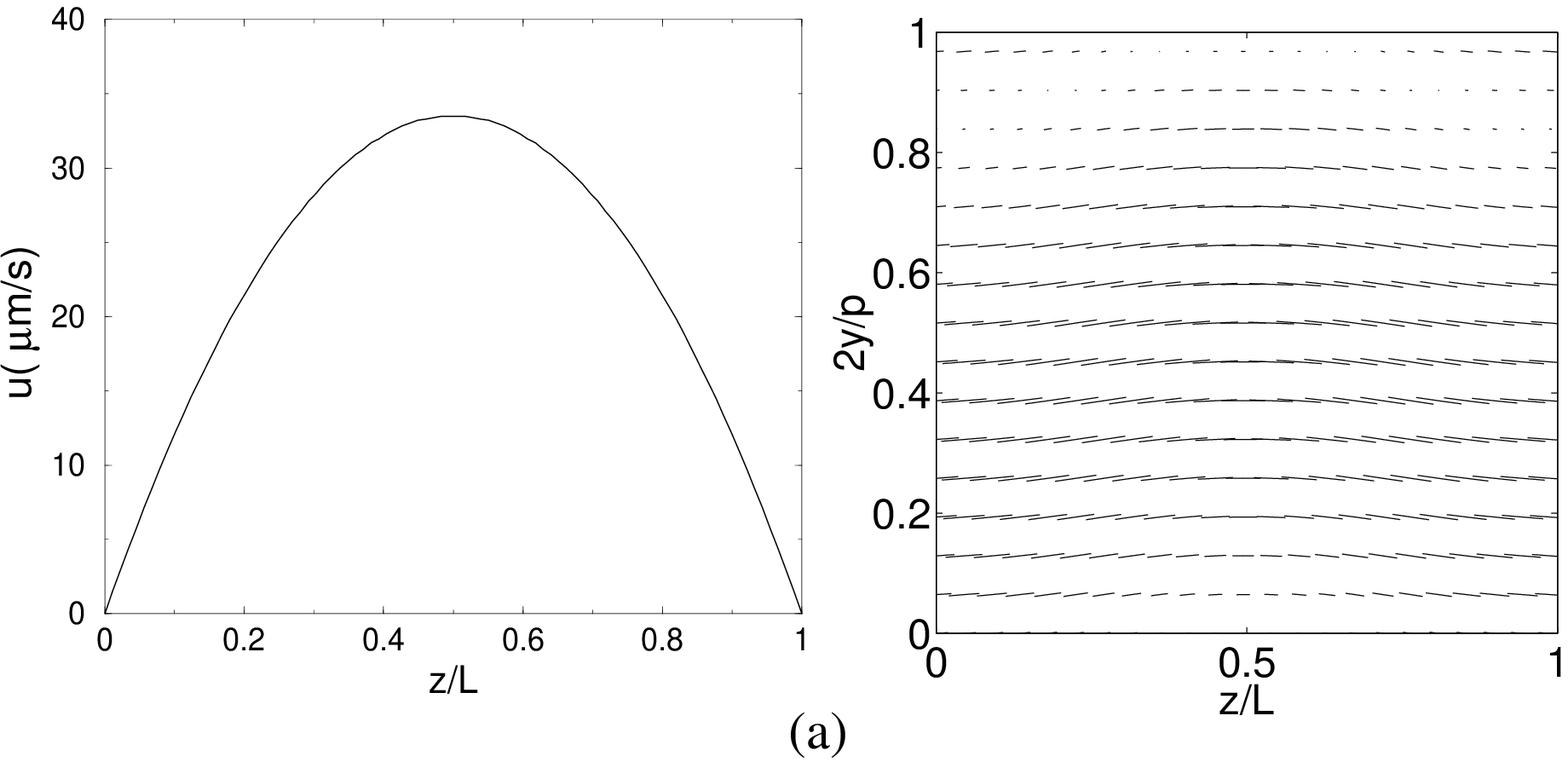,width=3.5in}}
\centerline{\psfig{figure=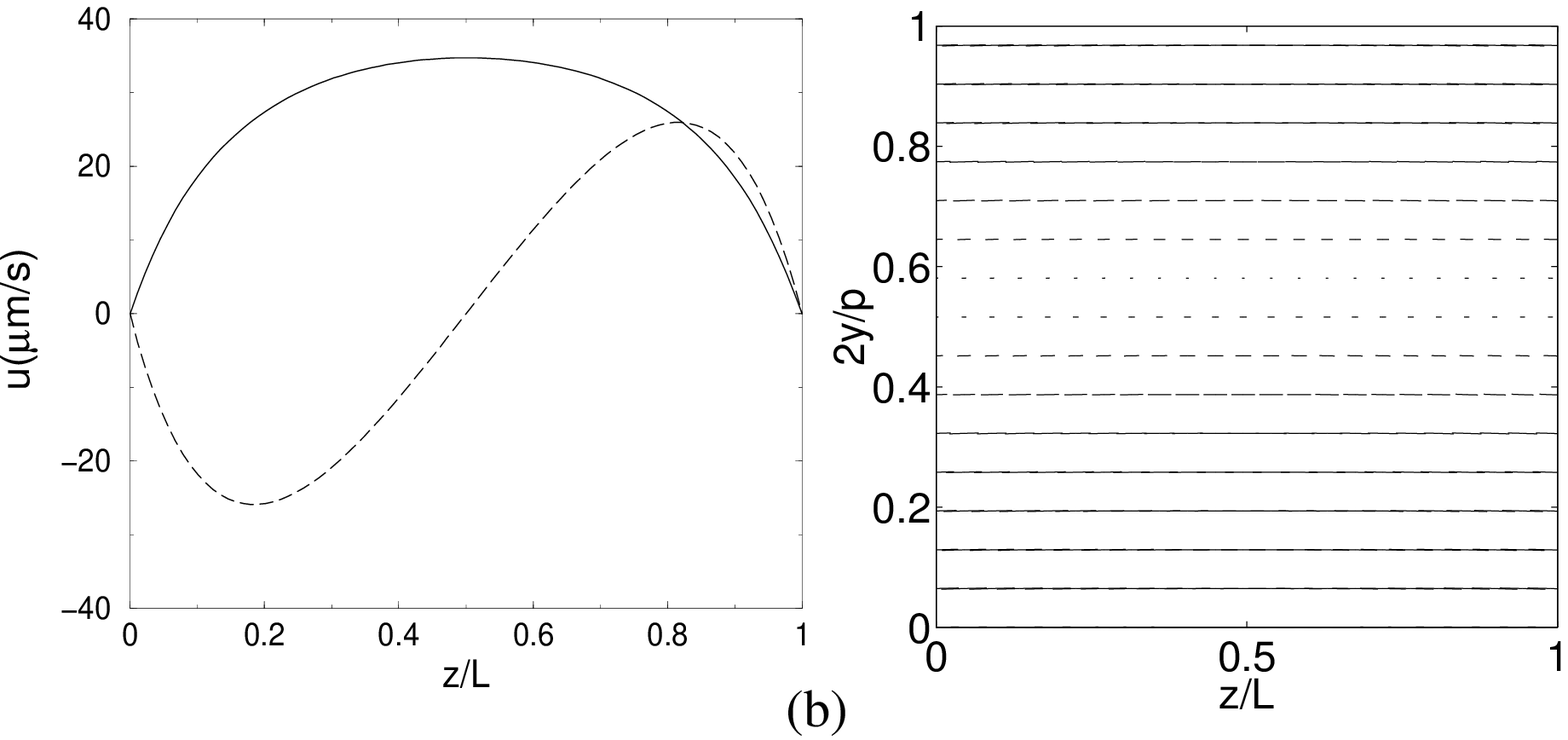,width=3.5in}}
\centerline{\psfig{figure=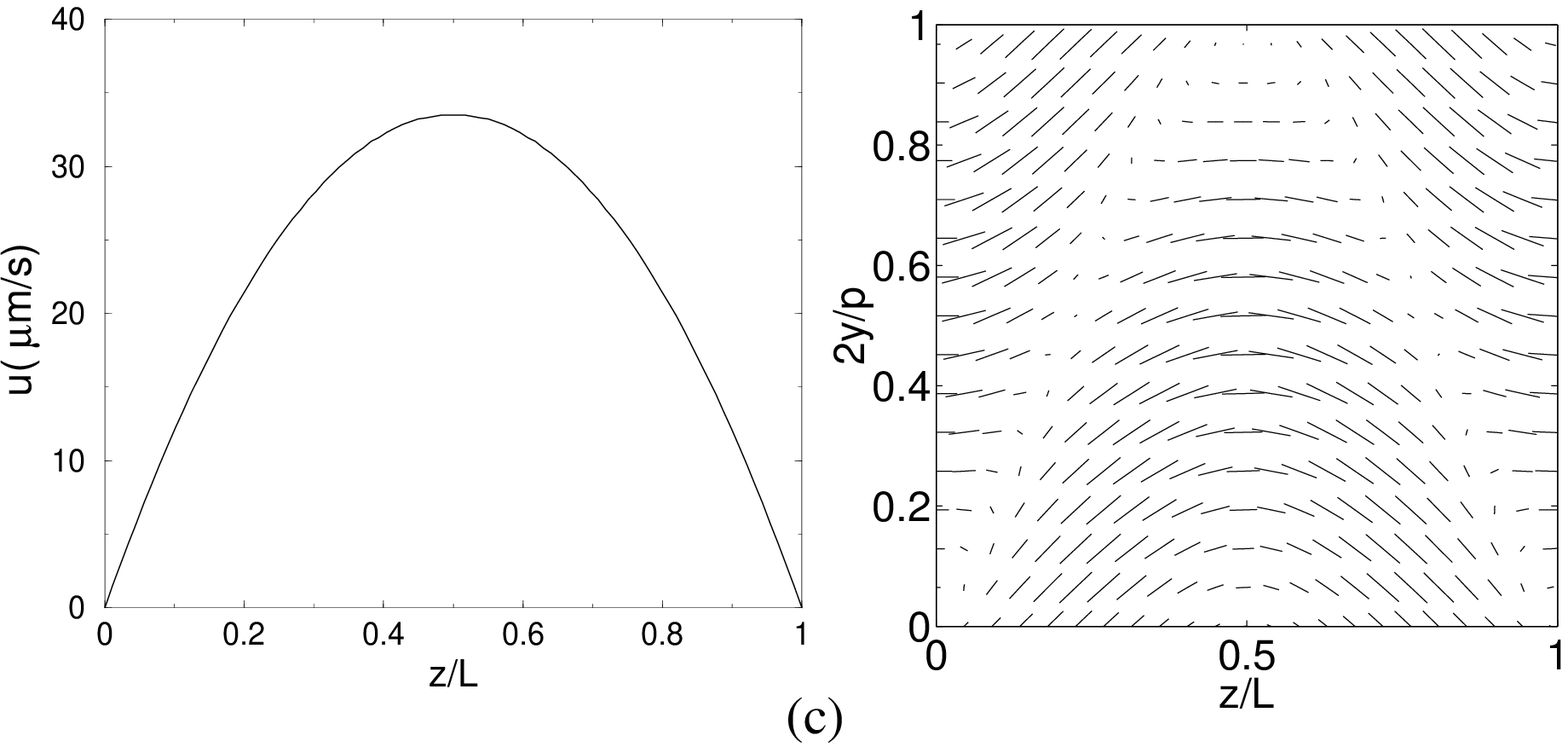,width=3.5in}}
\centerline{\psfig{figure=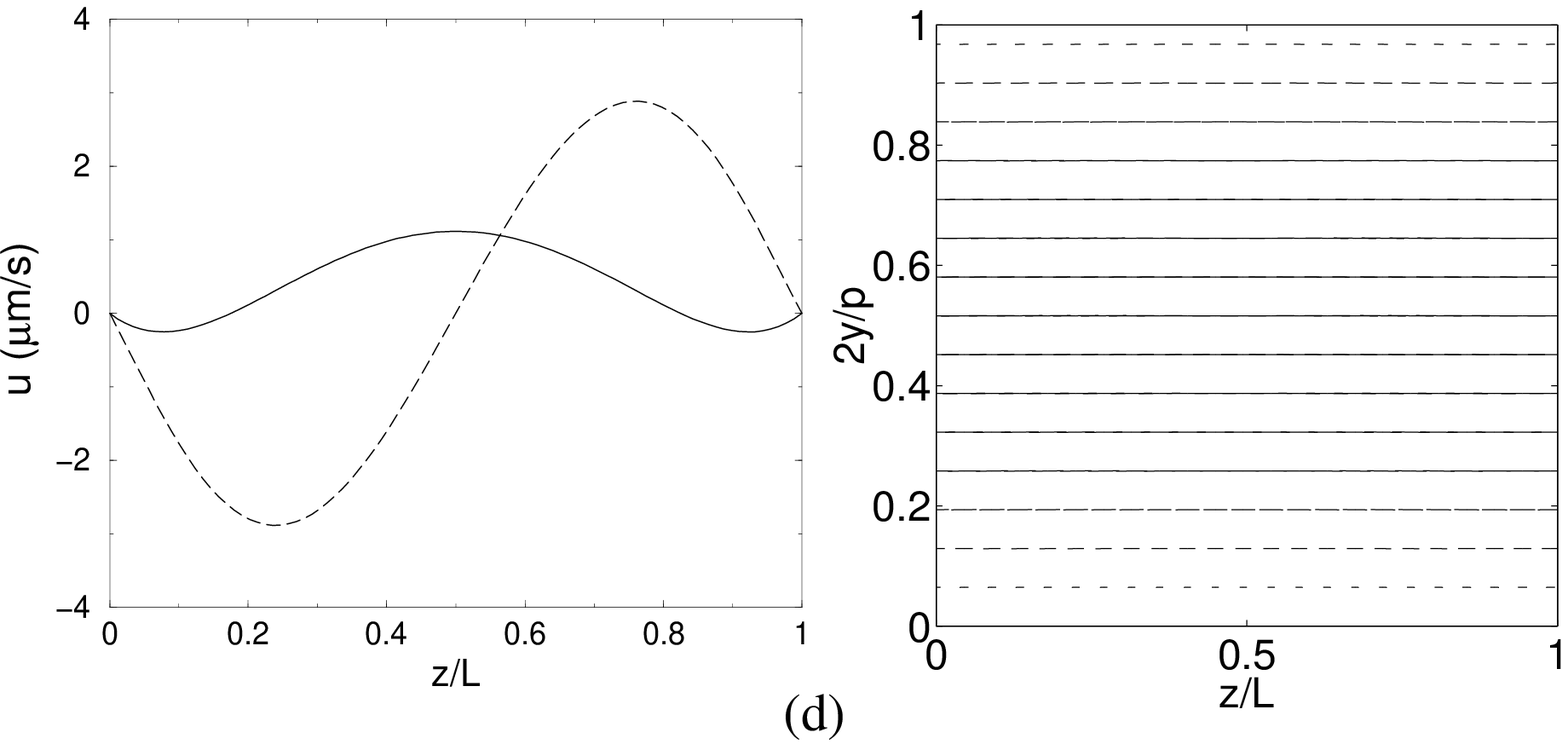,width=3.5in}}
\narrowtext
\caption{Velocity (left) and director fields (right) 
for (a) free boundary conditions,
no back-flow; (b) free boundaries and back-flow; (c) fixed boundaries, no
back-flow; (d) fixed boundaries and back-flow.
In (b) and (d) we also show the secondary flow (dashed line). 
This is an averaged velocity because the velocity field changes 
slightly along the flow direction $y$ due to the
helical arrangement of the director field.
The system studied corresponds to a cell of thickness $L=2.25$ $\mu$m,
and to a cholesteric liquid crystal with $p=1.6$ $\mu$m, $K\sim 6.3$ pN. 
The rotational viscosity is $\gamma_1=1$ Pois, 
while the ratio between the Leslie viscosities $\alpha_2$ and $\alpha_3$ 
is $\alpha_3/\alpha_2\sim -0.23$ (flow tumbling). }
\end{figure}

Figure 1b compares the case when the full equations of motion,
including back-flow, are solved. The velocity is 
similar
in magnitude, and the parabola flattens. This effect  becomes more
pronounced with an increase in the system size (Fig. 2). Our results are
consistent with the picture described in De Gennes and 
Prost\cite{degennes} (although
these authors consider pinned director boundary conditions) that the
velocity profile in a wide system will be flat, decaying to zero near
the boundaries in a length of order the pitch of the
cholesteric helix to satisfy the
no-slip conditions on the velocity. Also in this case 
the cholesteric helix drifts with the flow. 
A significant effect of the back-flow is that there is 
much less bending of the helix, the pitch remaining constant across the
system. 

\begin{figure}
\centerline{\psfig{figure=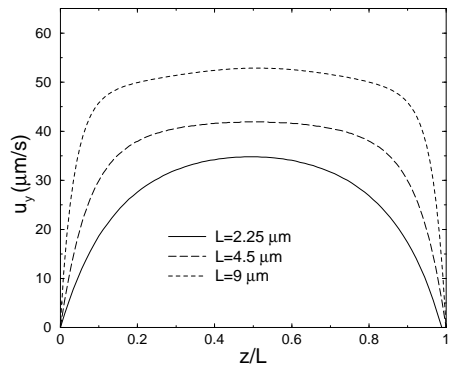,angle=270,width=2.3in} }
\caption{Velocity profiles as a function of scaled distance along
z. The pressure difference is scaled so that the velocity field of 
the liquid crystal in the isotropic phase is the same for all 
three sizes.
The other parameters are as in Fig. 1.}
\end{figure}
 
In the steady state we also find a secondary flow, of odd parity 
in $z$, along the $x$ direction, which attains its maxima (in 
absolute value) close to the boundaries, and is zero in the center
of the channel. The magnitude of this flow is comparable to the maximum
fluid velocity along $y$ and it gives rise
to shear forces which allow Eq. \ref{Qevolution} to be balanced
with a smaller director field deformation.

Now consider the case where the director is pinned at the wall 
(in such a way that there is no frustration in the helix when the 
system is at rest). Figure 1c shows the case without back-flow. The velocity
field is, as expected, identical to that in Figure 1a because the
director is having no effect on the flow. The helical texture is however
much more deformed with a substantial component along $y$. 
This occurs because the director configuration is unable to drift 
with the flow and thus the term $\vec u\cdot\vec{\nabla}{\bf Q}$
(in Eq. (\ref{Qevolution})) requires a larger balancing molecular field.

When back-flow is turned on for this case there is a striking
difference in the velocity profile. The net velocity is 
zero, to within the accuracy of the simulations 
(as would be expected from approximate analytic treatments
\cite{degennes,helfrich}.)
Secondary flow is important: it is bigger than the maximum velocity
attained in the primary flow. Because the velocities in the system are
very small the director field remains very close to its zero-flow
configuration.

We now consider what happens as the pressure difference is increased. 
First the chevrons gradually 
become more bent until a
threshold forcing at which a new structure appears.  This structure,
shown in Figure 3, has a flow-induced twist in the $z$-direction. 
(The solution pictured in Figure 1c corresponds to the threshold
between the chevron and the double twisted texture.) The
period of the twist along $z$ is roughly equal to the natural twist of
the cholesteric liquid crystal.  This double twist structure was found
for both flow-aligning (typically found in display devices) 
and flow tumbling (used in some rheological experiments\cite{experiments3}) 
viscosity regimes.

\begin{figure}
\centerline{\psfig{figure=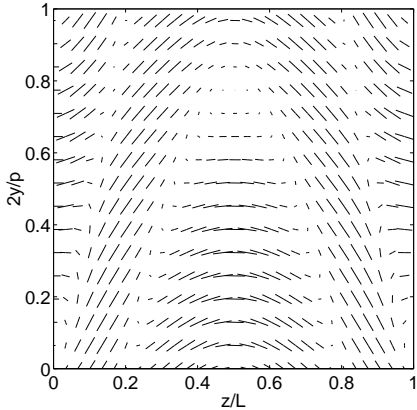,width=2.2in}}
\narrowtext
\caption{Director field for a system with
fixed boundary conditions. Parameters are $L=3$ $\mu$m, $p=1.6$ $\mu$m,
$K\sim 6.3$ pN, $\alpha_3/\alpha_2\sim 0.08$ (flow aligning). The maximum
velocity in equilibrium is $\sim 45$ $\mu$m s$^{-1}$ (for $\gamma_1=1$ Pois).
Note the double twist texture
near the center of the channel.}
\end{figure}

The maximum fluid velocity beyond which double twist appears depends
sensitively on the system parameters.  It is consistently smaller if
fixed boundary conditions are used.  For example, for 
$K\sim 6$ pN, $L \sim 2.25 \mu m$, $\alpha_3/\alpha_2\sim -0.23$ 
and $p\sim 1.6$ $\mu$ m, 
with fixed boundary conditions the velocity 
threshold is $\sim 30 \mu m s^{-1}$ while for free boundaries it is 
an order of magnitude bigger.  
As $L$ increases, the threshold velocity decreases
if fixed boundaries are used. This effect is somewhat
less pronounced with free boundaries. The crossover
between chevrons and double twist
is smooth; even at weak forcing one can identify a small twist
deformation in the $z$ direction.

It is possible to explain qualitatively the appearance of
a double twist by again looking at Eq. \ref{Qevolution}. 
Let us consider the case of anchored boundaries.
Near the center of the channel, the
solution is still determined by balancing the molecular field
with $\vec{u}\cdot\vec{\nabla}{\bf Q}$,
but since the flow is faster, the bending of the chevrons
become progressively more enhanced. When the 
deformation is such that two regions 
ordered along $x$ are a distance $\sim p$ apart 
along the $z$ axis, a doubly twisted state is expected to be more stable.
Such a texture is reminiscent of the doubly twisted cylinders which are 
the basic  constituents of blue phases\cite{degennes}. 
Indeed, in the case of no back-flow, where the simulations are stable  
for faster flows, a stack of such doubly twisted cylinders, separated by a 
regular array of defects, is stabilized by a large enough forcing.

Our results suggest that in real 
experiments\cite{experiments1,experiments2,experiments3}, 
for slow flows, the helix is likely pinned to 
its initial configuration (e.g. by irregularities), so that the situation 
is close to that in Figure 1d and the viscosity is very large. If the 
pressure difference is increased, either the pinning is destroyed 
or a double twist forms. In both cases the
viscosity would decrease as observed
experimentally. 

In conclusion, we have presented lattice Boltzmann simulations 
able to successfully simulate Poiseuille flow in cholesteric
liquid crystals in the permeation mode. For weak forcing, 
if the cholesteric helix is pinned at the boundaries, we
find a remarkable increase in the viscosity. With free boundaries there is 
a plug-like velocity profile in which the helix drifts with the flow. 
Beyond a threshold, dependent
on system parameters, the flow distorts the helix much more 
giving rise to a doubly twisted director pattern. 
This approach could also be used to investigate, for the first time, 
the rheology of blue phases. Predictions can tested with
present day microchannel experiments. In particular
a fast Poiseuille flow may provide a novel method of
inducing double twist in a liquid crystal texture thus allowing controlled
experiments on blue phases in cholesterics.

This work was supported by EPSRC grant no. GR/R83712/01.

\end{multicols}

\end{document}